\documentclass[10pt,letterpaper]{article}
\usepackage[top=0.85in,left=2.75in,footskip=0.75in]{geometry}

\usepackage{amsmath,amssymb}

\usepackage{changepage}

\usepackage[utf8x]{inputenc}

\usepackage{textcomp,marvosym}

\usepackage{cite}

\usepackage{nameref,hyperref}

\usepackage{microtype}
\DisableLigatures[f]{encoding = *, family = * }

\usepackage[table]{xcolor}

\usepackage{array}

\usepackage{mathtools}
\usepackage{physics}
\usepackage{mathptmx}
\usepackage{dutchcal}
\usepackage{ragged2e}

\usepackage{xcolor}
\usepackage{soul}
\usepackage{algorithm}
\usepackage{algpseudocode}
\usepackage{tabularx}
\usepackage{makecell}

\newcolumntype{+}{!{\vrule width 2pt}}

\newlength\savedwidth

\raggedright
\usepackage[aboveskip=1pt,labelfont=bf,labelsep=period,justification=raggedright,singlelinecheck=off]{caption}

\makeatletter
\renewcommand{\@biblabel}[1]{\quad#1.}
\makeatother

\usepackage{lastpage,fancyhdr,graphicx}
\usepackage{epstopdf}
\fancyheadoffset[L]{2.25in}
\fancyfootoffset[L]{2.25in}
\begin{document}
\vspace*{0.2in}

\begin{flushleft}
{\Large
\textbf\newline{L0-regularized compressed sensing with Mean-field Coherent Ising Machines} 
}
\newline
\\
Mastiyage Don Sudeera Hasaranga Gunathilaka \textsuperscript{1*},
Yoshitaka Inui\textsuperscript{2},
Satoshi Kako \textsuperscript{2},
Kazushi Mimura \textsuperscript{{1,3}},
Masato Okada \textsuperscript{4},
Yoshihisa Yamamoto\textsuperscript{2,5},
Toru Aonishi \textsuperscript{{1,4}}
\\
\bigskip
\textbf{1} School of Computing, Tokyo Institute of Technology, Tokyo, Japan
\\
\textbf{2} Physics and Informatics Laboratories, NTT Research Inc.,
940 Stewart Dr, Sunnyvale, CA 94085, USA
\\
\textbf{3} Graduate School of Information Sciences, Hiroshima City University, Hiroshima, Japan
\\
\textbf{4} Graduate School of Frontier Sciences, The University of Tokyo, Kashiwa, Chiba, Japan{ }
\\
\textbf{5} E. L. Ginzton Laboratory, Stanford University, Stanford, CA 94305, USA
\\
\bigskip

%
%





* mastiyage.s.aa@m.titech.ac.jp

\end{flushleft}
\section*{Abstract}
Coherent Ising Machine (CIM) is a network of optical parametric oscillators that solves combinatorial optimization problems by finding the ground state of an Ising Hamiltonian. As a practical application of CIM, Aonishi {\textit{et al}.} proposed a quantum-classical hybrid system to solve optimization problems of L0-regularization-based compressed sensing (L0RBCS). Gunathilaka {\textit{et al}.} has further enhanced the accuracy of the system. However, the computationally expensive CIM{'s} stochastic differential equations (SDEs) limit the use of digital hardware implementations. As an alternative to Gunathilaka \textit{et al}{.}'s CIM SDEs used previously, we propose using the mean-field CIM (MF-CIM) model, which is a physics-inspired heuristic solver without quantum noise. MF-CIM surmounts the high computational cost due to the simple nature of the differential equations (DEs). Furthermore, our results indicate that the proposed model has similar performance to physically accurate SDEs in both artificial and magnetic resonance imaging data, paving the way for implementing CIM-based L0RBCS on digital hardware such as Field Programmable Gate Arrays (FPGAs).


\section*{Introduction}
Over the past few years, quantum-inspired Ising machines have become increasingly popular \cite{qc1super,qc2super, cmososci1, cmososci2, nanoosci1, nanoosci2, nanoosci3, memrister1, memrister2, memrister3, cim2, unconArchs3}. 
{These Ising machines have demonstrated proficient computational capability in solving combinatorial optimization problems (COPs), especially quadratic unconstrained binary optimization (QUBO), which is ubiquitous in practical optimization problems.} A few examples of such problems include scheduling \cite{schedulling1, schedulling2}, portfolio optimization \cite{ising_portofolio}, and drug discovery \cite{ising_drug}.
In order to solve a COP mapped to an {Ising optimization problem, one must find the ground state of the Ising Hamiltonian.} By doing so, the optimal answer to the problem can be found. However, in practice, finding the ground state of an Ising Hamiltonian is rather complicated. In terms of complexity, this problem belongs to the class of nondeterministic polynomial-time (NP-hard) problems \cite{isingnp}.

{There have been several advances in the development of quantum-inspired Ising machines throughout the world} \cite{dwave2008, Yamamoto2, goto1,poormans2019}. {Among the array of architectures for Ising machines, the Coherent Ising Machine (CIM) stands as a notable contender. Employing optical parametric oscillators (OPOs), with the emergence of strong oscillatory states transcending predefined thresholds, CIM amplitude configurations effectively encode optimal solutions for specific Ising problems} \cite{Yamamoto2,cimqubo1}. {Despite the demonstrated capability of physical CIMs available in laboratory settings to accommodate 100,000 degenerate OPO pulses, equating to 100,000 spins} \cite{unconArchs3} {, persistent challenges endure. Notably, CIMs exhibit deficiencies in the precision controllability of physical parameters and stability} \cite{unconArchs3}. {Additionally, their mutual coupling is currently restricted to low-bit representations such as (-1, +1) to enable fast measurement feedback. These challenges are prevalent among nearly all physical architectures. As a practical remedy to these issues, the implementation of optimization algorithms rooted in classically computable models into digital hardware platforms such as FPGAs is being actively pursued.} {The rationale behind the pursuit of digital hardware implementation lies in the inherent stability, reliability, and cost-effectiveness of such systems, in contrast to their physical counterparts. Ref. } \cite{sbm100000} {indicates that FPGA clusters facilitating the implementation of Simulated Bifurcation Machines (SBMs) can potentially accommodate over 100,000 spins}. {However, the primary impediment to digital hardware implementation resides in the complexity of such optimization algorithms.} {Deriving classically computable models for CIMs entails the approximation of density matrix master equations utilizing quasi-probability distribution functions. {In the derivation of classically computable models for CIMs, the truncated-Wigner and the Positive-$P$ representations {have been} used.} \cite{Inui, Inui2022}

{Expanding the density matrix master equations using either the truncated-Wigner or Positive-$P$ representation facilitates the derivation of Langevin equations for the CIM. Nonetheless, numerical simulations of the derived stochastic differential equations (SDEs) are intricate and computationally demanding, rendering them unsuitable for large-scale simulations and implementation into dedicated digital hardware platforms such as FPGAs and Application-Specific Integrated Circuits (ASICs).} {There is, however, a simplified set of differential equations (DEs) called the mean-field CIM model, which is described in Refs.} \cite{Leleu1, AmpLeleu, ng, gunathilakameanfield}, {offers an alternative. In these equations, quantum noise and measurement effects are not considered. While this model deviates from the precise representation of the physical CIM, it offers a simplified framework. This simplification enables various numerical analyses, such as linear stability analysis, and facilitates the development of simple heuristic algorithms suitable for large-scale simulations and implementation on dedicated digital hardware. Recently, the mean-field CIM model has been extended to incorporate the Zeeman term using an effective implementation technique known as chaotic amplitude control (CAC), as detailed in Ref.} \cite{gunathilakameanfield}. {Consequently, the mean-field CIM model, augmented with the Zeeman term via CAC, finds applications in diverse large-scale real-world COPs, including Code Division Multiple Access (CDMA), Multi-Input Multi-Output (MIMO), and L0-regularization-based Compressed Sensing (L0RBCS)} \cite{Aonishi, gunathilaka2023}. {However, to date, investigations using this model have primarily focused on small-scale Sherrington-Kirkpatrick (SK) problems.}

{In this paper, the mean-field CIM model with the Zeeman term incorporated by CAC is applied to L0RBCS, which is a COP of image reconstruction. We investigate how the mean-field CIM model can replace physically accurate CIM models such as Positive-{$P$} SDEs for realizing L0RBCS. Our results demonstrate that both the mean-field CIM model and the positive-$P$ CIM model perform similarly, despite the fact that the computational costs associated with the mean-field CIM model are lower. We describe in another paper the implementation of large-scale optimization algorithms based on the mean-field CIM model into FPGAs as well as its fast computation of L0RBCS} \cite{aonishifpga}.

\section*{Methods}

\subsection*{Alternative Minimization Algorithm for L0RBCS}

The compressed sensing (CS) method involves the reconstruction of high-dimensional signals or images from highly downsampled measurements {achieved through the following optimisation.}

\begin{equation}
\label{l0init2}
    \hat{x} = \operatorname*{argmin}_{x \in \mathbb{R}^N}\|x\|_{p} \ \ s.t \ y = Ax .
\end{equation}

\noindent In the eq. (\ref{l0init2}), an observed signal $y \in \mathbb{R}^M$, an observation matrix $A \in \mathbb{R}^{M\times N}$, and a source signal $x \in \mathbb{R}^N$ {are indicated}. 
The sparseness of the vector $x$, which corresponds to the {ratio of} of non-zero elements in the vector is indicated as $a$. {Compression ratio refers to the ratio of the number of rows to the number of columns of matrix $A$, denoted by $\alpha$.}

An attempt at solving L0-norm CS, which is a COP, can be achieved by considering the two-fold formulation of eq. (\ref{l0init2}) \cite{twofold1,twofold2}.

\begin{equation}
\label{l0}
    (\hat{R}, \hat{\sigma}) = \operatorname*{argmin}_{\sigma \in \{0,1\}^{N}}\operatorname*{argmin}_{R\in\mathbb{R}^{N}} \left(\| y - A(\sigma \circ R)\|_{2}^{2}\right) \ \ s.t \   \|\sigma\|_{0} \le \Omega .
\end{equation}

\noindent Here $R \in \mathbb{R}^N$ and $\sigma \in \left\{{0,1}\right\}^N$ correspond to the source signal and support vector, respectively. The support vector consists only of binary values. A zero entry indicates that the source signal entry is zero, and a non-zero entry suggests that the source signal entry is non-zero. The condition $\|\sigma\|_{0} \le \Omega$ is a sparsity-inducing prior for constraining the number of non-zero entries to be $\Omega$.

Ref. \cite{Aonishi} and Ref. \cite{gunathilaka2023} employed a quantum-classical hybrid approach to solve optimization problems of L0-norm CS.
In order to implement the L0-norm CS with the hybrid system, the following regularization form {called L0RBCS} is used \cite{Aonishi, gunathilaka2023}.

\begin{equation}
\label{doublel0}
    (R, \sigma) = \operatorname*{argmin}_{\sigma \in \{0,1\}^{N}}\operatorname*{argmin}_{R\in\mathbb{R}^{N}} \left(\frac{1}{2} \| y - A(\sigma \circ R)\|_{2}^{2} + {\lambda} \|\sigma\|_{0}\right) .
\end{equation}

\noindent The element-wise representation of eq. (\ref{doublel0}) gives the following Hamiltonian.

\begin{equation}
\label{l0Hamiltonian}
    \mathbcal{H} = \sum_{r<r'}^{N}\sum_{k = 1}^{M} A_{r}^{k}A_{r'}^{k}R_{r}R_{r'}\sigma_{r}\sigma_{r'} - \sum_{r=1}^{N}\sum_{k =1}^{M} y^{k}A_{r}^{k}R_{r}\sigma_{r} + {\lambda} \sum_{r = 1}^{N} \sigma_r , 
\end{equation}

\noindent where an element $A^k$ in $A$, an element $y^k$ in $y$, an element $R_r$ in $R$ and an element $\sigma_r$ in $\sigma$. 
The problem of optimizing with regards to $\sigma$ is a quadratic unconstrained binary optimization (QUBO). This hybrid approach optimizes $\sigma$ with the use of a CIM while optimizing $R$ with a Classical Digital Processor (CDP). 

\subsection*{Closed-loop CIM for L0RBCS}\label{CLCIMSDE}

{To date, two proposals have been made for solving L0RBCS using CIMs: an open-loop CIM model with CDP {(OL-CIM-CDP)} developed by Aonishi {\textit{et al}.}} \cite{Aonishi} {and a closed-loop CIM model with CDP {(CAC-CIM-CDP)} developed by Gunathilaka {\textit{et al}.}} \cite{gunathilaka2023}. {The closed-loop CIM model integrates additional nonlinear dynamical feedback mechanisms to actively regulate system amplitudes towards a designated value known as the target amplitude. This implementation induces chaotic dynamics within the system and potentially enables escape from local minima} \cite{Inui2022, AmpLeleu, Leleu1, gunathilaka2023}. {Conversely, the open-loop CIM model is devoid of such nonlinear dynamical feedback. In CAC-CIM-CDP in Ref. }\cite{gunathilaka2023}, {there were two models based on the truncated-Wigner and Positive-$P$ representations. It was shown that both models perform almost identically in low-quantum noise situations, and outperforms OL-CIM-CDP} \cite{Inui2022,gunathilaka2023}. {Thus, this paper only considers the Positive-{$P$} model, which we will refer to as CAC-CIM-CDP (Positive-$P$).}

{For L0RBCS, the SDEs normalized by $g$ for CAC-CIM-CDP (Positive-$P$) are described as follows.}

\begin{equation}
\label{ppGACsCIM1}
\begin{multlined}
        \dfrac{d}{dt}\mu_{r} = - \left(1 -p + j\right)\mu_{r} - {\mu_{r}\left(\mu_{r}^{2} + 2g^2n_r + g^2m_r\right)} \\  + \sqrt{jg^2}\left(m_r + n_r\right)W_{R,r}+ {{K}}\left(\frac{d\mu_{r}}{dt}\right)_{inj,r} ,
\end{multlined}
\end{equation}

\begin{equation}
\label{ppGACsCIM2}
\begin{multlined}
        {\dfrac{d}{dt}n_{r} = -2 \left(1 + j\right)n_r + 2pm_r - 2\mu_{r}^2\left(2n_r + m_r\right)} {- j\left(m_r + n_r\right)^{2}},
\end{multlined}
\end{equation}
\begin{equation}
\label{ppGACsCIM3}
\begin{multlined}
        {\dfrac{d}{dt}m_{r} = -2 \left(1 + j\right)m_r + 2p n_r - 2\mu_{r}^{2}\left(2m_r + n_r\right) + } p\\ -  \left(\mu_{r}^{2} + g^{2}m_r\right) - j\left(m_r + n_r\right)^{2} .
\end{multlined}
\end{equation}

\noindent {The variable} $\mu_r$ represents the normalized mean-amplitude of the $r$-th DOPO pulse {($\mu_r = g\langle\hat{a}_r\rangle$)}, whereas $m_r = \langle\delta\hat{a}^2_r\rangle$ and $n_r = \langle\delta\hat{a}^{\dagger}_r\delta\hat{a}_r\rangle$ correspond to the variances of quantum fluctuations ({Here $\hat{a}_r$ indicates the annihilation operator of the $r$-th signal mode and $\delta\hat{a}_r = \hat{a}_r - \langle\hat{a}_r\rangle$}).
{$W_{R,r} (t)$ indicates i.i.d. real Gaussian noise} satisfying $\langle W_{R,r} (t)\rangle =0$ and $\langle W_{R,r} (t)W_{R,r'} (t')\rangle =\delta_{rr'}\delta(t-t')$. {The coefficient} $g^2$ is the nonlinear saturation parameter of the CIM. $p$ and ${K}$ indicate the pump rate and the feedback strength. {The parameter} $j$ is the {total} normalized out-coupling rate {at two beam splitters including one} for optical homodyne measurement. 
The linear loss is represented by $(1+j)$ in eq. (\ref{ppGACsCIM1}), while the two-photon loss is represented by $\mu_{r}\left(\mu_{r}^{2} + 2g^2n_r + g^2m_r\right)$. The term $\sqrt{jg^2}\left(m_r + n_r\right)W_{R,r}$ represents the mean-amplitude shift caused by quantum measurements. For this model, injection field $(d\mu_r/dt)_{inj,r}$ is as follows.

\begin{equation}
\label{GACSlocalfieldmain}
    \left(\dfrac{d\mu_{r}}{dt}\right)_{inj,r} = je_r\left( R_rh_r - \dfrac{\sqrt{{\tau}}\eta^2}{4}\right),
\end{equation}
\begin{equation}
\centering
\label{GaussianEC5}
        {\dfrac{d}{dt}e_{r} = -\beta\left(\tilde{\mu}_{r}^2 - \tau\right)e_{r}},
\end{equation}
\begin{equation}
\label{GACsCIM3}
        \tilde{\mu}_{r} = \mu_{r} + \sqrt{\frac{g^2}{4j}}W_{R,r},
\end{equation}
\begin{equation}
\centering
\label{localfieldGACS}
    h_r = -{\sum_{r' = 1 (\neq r)}^{N}\sum_{k = 1}^{M}} A_{r}^{k}A_{r'}^{k}R_{r'}\dfrac{1}{2}\left(\tilde{\mu}_{r'} + \sqrt{{\tau}} \right) {+} \sum_{k = 1}^{M} \sqrt{{\tau}}{A_{r}^{k}y^{k}},
\end{equation}

\noindent where $h_r$ denotes the local field, and $e_r$ is the auxiliary variable for error feedback. The target amplitude for CAC is given by {the parameter} $\tau$. {The variable} $R_r$ is the signal value estimated by CDP. {$\eta$ represents the threshold given by $\eta = \sqrt{2\lambda}$, which was introduced to maintain consistency with previous research, and represents the threshold value of the L0-norm proximal operator (hard thresholding)} \cite{Aonishi, gunathilaka2023}. $\tilde{\mu}_r$ implies the measured-amplitude, and $W_{R,r}$ is the {same} real Gaussian noise as used in eq. (\ref{ppGACsCIM1}). The mutual interaction can be considered as $\tilde{J}_{rr'} = -\sum_{k = 1}^M A_r^k A_{r'}^k$ and the Zeeman term is $h_r^z = \sqrt{{\tau}}\sum_{k=1}^M A_r^k y^k$. Refer to Ref. \cite{gunathilaka2023} for a detailed explanation. Closed-loop {and} open-loop systems differ in whether the injection field of the CIM system is dynamically modulated or not \cite{ng,gunathilakameanfield}.

\subsection*{Closed-loop Mean-field CIM for L0RBCS} \label{MFZSDE}

Given the previous section, it is obvious that the SDEs used for simulation are quite complicated and pose a high computational cost. 

 Our approach in this paper relies on the physics-inspired heuristic CIM algorithm without quantum noise and disregards any measurement effects proposed in Ref.\cite{gunathilakameanfield}, called mean-field CIM DEs with CAC to overcome this computational cost problem. Hereafter, mean-field CIM with CAC will be referred to as MFZ-CIM {(Mean-Field Zeeman CIM)} and mean-field CIM with CAC and CDP as CAC-MFZ-CDP.
 {By considering $\mu_r=c_r$, ignoring the fluctuations and taking the limit as $g^2 \rightarrow 0$, as described in Ref.} \cite{gunathilakameanfield}, {eqs.} (\ref{ppGACsCIM1})-(\ref{ppGACsCIM3}) {can be simplified as follows} \cite{gunathilaka2023,gunathilakameanfield}.
 
\begin{equation}
\centering
\label{mfeq1}
        \frac{dc_r}{dt} = (-1 + p - c_{r}^2)c_r + K\left(\dfrac{dc_{r}}{dt}\right)_{inj,r} ,
\end{equation}

\begin{equation}
\label{MFZlocalfieldmain}
    \left(\dfrac{dc_{r}}{dt}\right)_{inj,r} = je_r\left( R_rh_r - \dfrac{\eta^2}{4}\sqrt{\tau}\right),
\end{equation}
\begin{equation}
\centering
\label{GMFZEC5}
        {\dfrac{d}{dt}e_{r} = -\beta\left(c_{r}^2 - \tau\right)e_{r}},
\end{equation}
\begin{equation}
\centering
\label{localfieldMFZ}
    h_r = -{\sum_{r' = 1 (\neq r)}^{N}\sum_{k = 1}^{M}} A_{r}^{k}A_{r'}^{k}R_{r'}\dfrac{1}{4}\left(c_{r'} + \sqrt{\tau} \right) {+} \sum_{k = 1}^{M} \dfrac{\sqrt{\tau}}{2}{A_{r}^{k}y^{k}}.
\end{equation}

\noindent On the right-hand side of the eq. (\ref{mfeq1}), linear loss, pump gain, and nonlinear saturation are represented by the first, second, and third terms {in the parenthesis}.
 $K\left({dc_{r}}/{dt}\right)_{inj,r}$ corresponds to the mutual coupling term. In eq.(\ref{MFZlocalfieldmain}) $h_r$ is the local field expressed as eq. (\ref{localfieldMFZ}), $e_r$ is for the error feedback in the CAC feedback loop, and $\tau$ indicates the target amplitude for the CAC. $R_r$ is the signal value estimated by the CDP. 
 
 This threshold $\eta$ is given by $\eta = \sqrt{2\lambda}$ so that consistency can be maintained with eq. (\ref{GACSlocalfieldmain}).
 In the local field eq. (\ref{localfieldMFZ}), the mutual interaction is  $\tilde{J}_{rr'} = -\sum_{k = 1}^M A_r^k A_{r'}^k$ and the Zeeman term is $h_r^z = \sqrt{{\tau}}\sum_{k=1}^M A_r^k y^k$.

The above local field eq.(\ref{GMFZEC5}) is the same as Ref.\cite{gunathilaka2023}. Next, consistent with Ref. \cite{Aonishi}, we introduce the local field wherein continuous amplitudes $1/2(c_r’ + \sqrt{\tau})$  are substituted with binary values $\sigma=H(c_r)$ ($H(x)$: Heaviside step function).

\begin{equation}
\centering
\label{localfieldMFZBN}
    h_r^{BN} = -\dfrac{\sqrt{\tau}}{2}\left({\sum_{r' = 1 (\neq r)}^{N}\sum_{k = 1}^{M}} A_{r}^{k}A_{r'}^{k}R_{r'}\sigma_{r'} {-} \sum_{k = 1}^{M} {A_{r}^{k}y^{k}}\right) .
\end{equation}

\noindent This is a simplified version of the local field (See supporting information for the derivation) {that makes} digital hardware implementations easier. At the initial state of eqs. (\ref{mfeq1})-(\ref{localfieldMFZ}), due to its deterministic nature, a Gaussian {random value} with {a mean of 0 and a} variance of $10^{-4}$ was introduced to $c_r$ as vacuum noise, {and $e_r$ is initialized as $1$.} 

\section*{Numerical experiments}\label{numexp}

We used artificial random data and Magnetic Resonance Image (MRI) data of a brain to evaluate the performance of the proposed model CAC-MFZ-CDP and its use of continuous ($h_r$) and binarized ($h_r^{BN}$) local fields.

{In accordance with Ref.} \cite{Aonishi,gunathilaka2023}, {we also employ the same observation model to generate random datasets.}

\begin{equation}
\label{observationmodelmatrix}
\begin{bmatrix}
y^{1} \\
y^{2} \\
\vdots\\
y^{M}
\end{bmatrix} = 
\begin{bmatrix}
A_{1}^{1} & A_{2}^{1} & \cdots & A_{N}^{1} \\
A_{1}^{2} & A_{2}^{2} & \cdots & A_{N}^{2} \\
\vdots  & \vdots  & \ddots & \vdots  \\
A_{1}^{M} & A_{2}^{M} & \cdots & A_{N}^{M}
\end{bmatrix}
\begin{bmatrix}
\xi_{1}x_{1} \\
\xi_{2}x_{2} \\
\vdots\\
\xi_{N}x_{N}
\end{bmatrix}
+
\begin{bmatrix}
w_{noise}^{1} \\
w_{noise}^{2} \\
\vdots\\
w_{noise}^{M}
\end{bmatrix} .
\end{equation}

\noindent Here a random observation matrix is generated from independent and identical normal distributions with the variance $1/M$, which meets the condition $\langle A_r^k\rangle =0$ and $\langle A_r^k A_{r'}^{k'}\rangle = 1/M \delta_{rr'} \delta_{kk'}$, for each entry of the observation matrix $A \in\mathbb{R}^{M\times N}$. Also, the true source signal $x\in\mathbb{R}^N$ is randomly generated from an independent and identical normal distribution with {a mean 0 and a} variance of $1$, which satisfies {$\langle x_r \rangle = 0$} and $\langle x_r x_{r'}\rangle = \delta_{rr'}$. Randomly selected $a\times N$ elements of $\xi\in(0,1)^N$ are assigned $1$, while others are assigned $0$. Refer to Ref. \cite{Aonishi,gunathilaka2023} for a detailed explanation. {{The variable} $w_{noise}\in\mathbb{R}^M$ indicates the observation noise satisfying $\langle w_{noise}^{k}\rangle =0$ and $\langle w_{noise}^{k} w_{noise}^{k'}\rangle =\nu^2 \delta_{kk'}$. $\nu^2$ is the variance of the observation noise.}

In order to evaluate the proposed approach on real-world data, an MRI of a brain from Ref. \cite{fastmri} was used. Using the BILINEAR interpolation method, the original $320\times 320$ brain MRI image was resized to $64\times 64$ and $128\times 128$ images. For the simulations, 78.8\% and 82.2\% of the Haar wavelet (HWT) coefficients were set to zero to produce two sparse images with different sizes ($64\times64$ and $128\times128$ pixels) spanned by Haar basis functions with sparseness of 0.212 and 0.178. 
Then after applying the discrete Fourier transform (DFT), 1638 and 4915 points were undersampled at random points from the $64\times64$ and $128\times128$ $k$-space data to produce two observation signals with compression rates of 0.4 and 0.3. We solved the following optimization problem based on our previous work to reconstruct the source signals from the undersampled $k$-space data.

\begin{equation}
\label{MRIl0init1}
    x = \operatorname{argmin}(\| y - SFx\|_{2}^{2} + \dfrac{1}{2}\gamma \|\Delta_{v}x\|_{2}^{2} +  \dfrac{1}{2}\gamma \|\Delta_{h}x\|_{2}^{2} + \lambda\|\Psi x\|_{0}) .
\end{equation}

\noindent In this case, $x$ is the source signal, while $y$ is the observation signal. DFT matrices are denoted by $F$, while HWT matrices are denoted by $\Psi$. Undersampling is performed at random points using an undersampling matrix, $S$. $\Delta_{v}$ and $\Delta_{h}$ are the matrices discretely representing the vertical and horizontal second-order derivative operators, respectively. $\gamma$ and $\lambda$ are the L2 and L0 regularization parameters. 

Considering HWT $r = \Psi x$ to eq. (\ref{MRIl0init1}), the mutual interaction matrix $J$ and the Zeeman term vector $h^z$ for CIM are given as follows.

\begin{equation}
\label{MRIzeeman}
    h^{z} = SF\Psi^{T}y,
\end{equation}
\begin{equation}
\label{MRIJMatrix1}
    \tilde{J} = \Psi F^{T}S^{T}SF\Psi^{T} + \gamma\Psi\Delta_{v}^{T}\Delta_{v}\Psi^{T} + \gamma\Psi\Delta_{h}^{T}\Delta_{h}\Psi^{T} .
\end{equation}

\noindent Here, the observation matrix is given as $A = SF\Psi^T$. The second and third terms in $\tilde{J}$ are from the L2 regularization terms. 
$\gamma$ is set to 0.0001. As an initial condition for the CIM simulation, we use the LASSO solution. Refer to Ref. \cite{Aonishi,gunathilaka2023} for a detailed explanation.

In this paper, CAC-MFZ-CDP is compared primarily with CAC-CIM-CDP (Positive-$P$) of Gunathilaka \textit{et al} \cite{gunathilaka2023}.
Depending on the time $t$, CAC-CIM-CDP and CAC-MFZ-CDP pump rates $p$ were scheduled as follows.

\begin{equation}
\label{pumprateGACS}
        p = (p_{thr} - d) + \frac{2d}{1+e^{-\left(\dfrac{t-4}{2}\right)}} .
\end{equation}

\noindent Here, $p_{thr} = 1$ for all simulations. A value of $d$ of 0.4 was used for artificial random data, while a value of 0.6 was used for MRI data simulations. In both CAC-CIM-CDP and CAC-MFZ-CDP, the threshold $\eta$ was scheduled depending on the alternating iteration time $i$ as follows.

\begin{equation}
\label{etaSchedulling}
        \eta_{i} = \max \left[\eta_{init} \left(1 - \dfrac{i}{velo}\right), \eta_{end}\right] .
\end{equation}

\noindent In all simulations using artificial random data, $velo$ is equal to $51$. And $velo = 11$ was used for CAC-MFZ-CDP and CAC-CIM-CDP in MRI simulations. In MRI data, the threshold $\eta$ was fixed by setting $\eta_{init} = \eta_{end}$ instead of linearly lowering as in eq. (\ref{etaSchedulling}).
Overall $g^2 = 10^{-7}$. In CAC-MFZ-CDP and CL-CIM-CDP, we set the time increment $\Delta t = 0.02$ increasing up to 20$\times$ the photon's lifetime. 

{For artificial random data, With the CDP, we used the Jacobi method with a time increment $\Delta t_c$ set to $0.1$ and $100$ iterations. With MRI data, the Conjugate Gradient Descent method was used for the CDP, with a maximum iteration of $10000$.} In the simulations with observation noise, for CAC-CIM-CDP we used $\tau = 0.21$ for $\nu = 0.05$, while $\tau = 0.15$ for $\nu = 0.1$ \cite{gunathilaka2023}. Throughout all simulations of CAC-MFZ-CDP, $\tau = 1$ was used.
Parameter $\eta_{init} = 0.8$ was used for CAC-CIM-CDP and CAC-MFZ-CDP. $\eta_{end}$ was set to 0.18 when $\nu = 0.05$ and 0.35 for $\nu = 0.1$ to keep consistency with previous research. $N = 2000$ was used in all artificial random data simulations. In the case of ${K}$, CAC-CIM-CDP (Positive-$P$) set {it} to 0.01 and for CAC-MFZ-CDP, it was set to 0.1 for MRI data. {In artificial random data} ${K} = 1$ for every model.

\begin{algorithm}
\caption{Alternating minimisation for L0RBCS as a QUBO problem on CAC-CIM-CDP and MFZ-CIM-CDP. The schedules of the pump rate and threshold are given in eq. (\ref{pumprateGACS}) and (\ref{etaSchedulling})}\label{algo}
\begin{algorithmic}[1]
\Require $M\times N$ observation matrix: $A,$ $M$-dimensional observation signal: $y$; 
\Ensure $N$-dimensional support vector: $\sigma$, $N$-dimensional signal vector: $R$;
\State Initialise $R = R_{init}$, $\eta = \eta_{init}$, $g^2 =10^{-7}$ for CAC-CIM-CDP (Positive-$P$), and $\tau = 1$
\For{$i = 0$ to $velo$}
    \State {Minimise $\mathbcal{H}$ with respect to $\sigma$ by CIM:}
    \State $\sigma$ = Support estimation using CAC-CIM (eq. (\ref{ppGACsCIM1})-(\ref{localfieldGACS})) or MFZ-CIM (eq. (\ref{mfeq1})-(\ref{localfieldMFZBN})) \newline
    \hspace*{2em} Initialise $c_r$ with random normal Gaussian values with mean 0 and variance $10^{-4}$ \hspace*{2em} and $e_r = 1$ for MFZ-CIM-CDP. {For CAC-CIM-CDP initialize $\mu_r = 0$, $n_r = 0$ and} \hspace*{2.3em} {$m_r = 0$.} And we increase {time to 20 times} the photon’s lifetime. \newline \hspace*{2em}
    
    \State Minimise $\mathbcal{H}$ with respect to $R$ by CDP using Conjugate Gradient Descent or Jacobi \hspace*{1.2em} method:\newline

    \State Update $\eta$
\EndFor
\end{algorithmic}
\end{algorithm}

\section*{Results}

\subsection*{Amplitude evolution of Continuous vs Binarized} \label{contvsbin}

Fig. \ref{amps} {illustrates the evolution of amplitudes $c_r$ in both MFZ-CIM DEs with binarized ($h_r^{BN}$) and continuous ($h_r$) local fields} (Fig. \ref{amps}b and Fig. \ref{amps}c respectively) {and the evolution of measured amplitudes $\tilde{\mu}_r$ in CAC-CIM SDEs (Positive-$P$)} (Fig. \ref{amps}a). {The indicated amplitudes for each model are from the second alternative minimization process in Algorithm} \ref{algo}. {CAC-CIM (Positive-$P$) and MFZ-CIM exhibit distinct amplitude evolutions despite solving the same problem instance, namely the CAC-CIM (Positive-$P$) is more chaotic than the MFZ-CIM. It is evident from these figures that, despite such distinct amplitude evolutions, the amplitudes for each model are equalized around $-1$ and $+1$ and are intermittently wandered between $-1$ and $+1$ as the target amplitude $\tau$ is set to $1$. The figures clearly demonstrate the oscillation between $-1$ and $+1$ as $\tau$ is set to $1$. A noteworthy observation is that, compared to MFZ-CIM with $h_r$, MFZ-CIM with $h_r^{BN}$ exhibits somewhat more chaotic behavior after the bifurcation, whilst MFZ-CIM with $h_r$ shows relatively more chaotic behavior near the bifurcation.}

\begin{figure}[!ht]
\includegraphics[width=135mm]{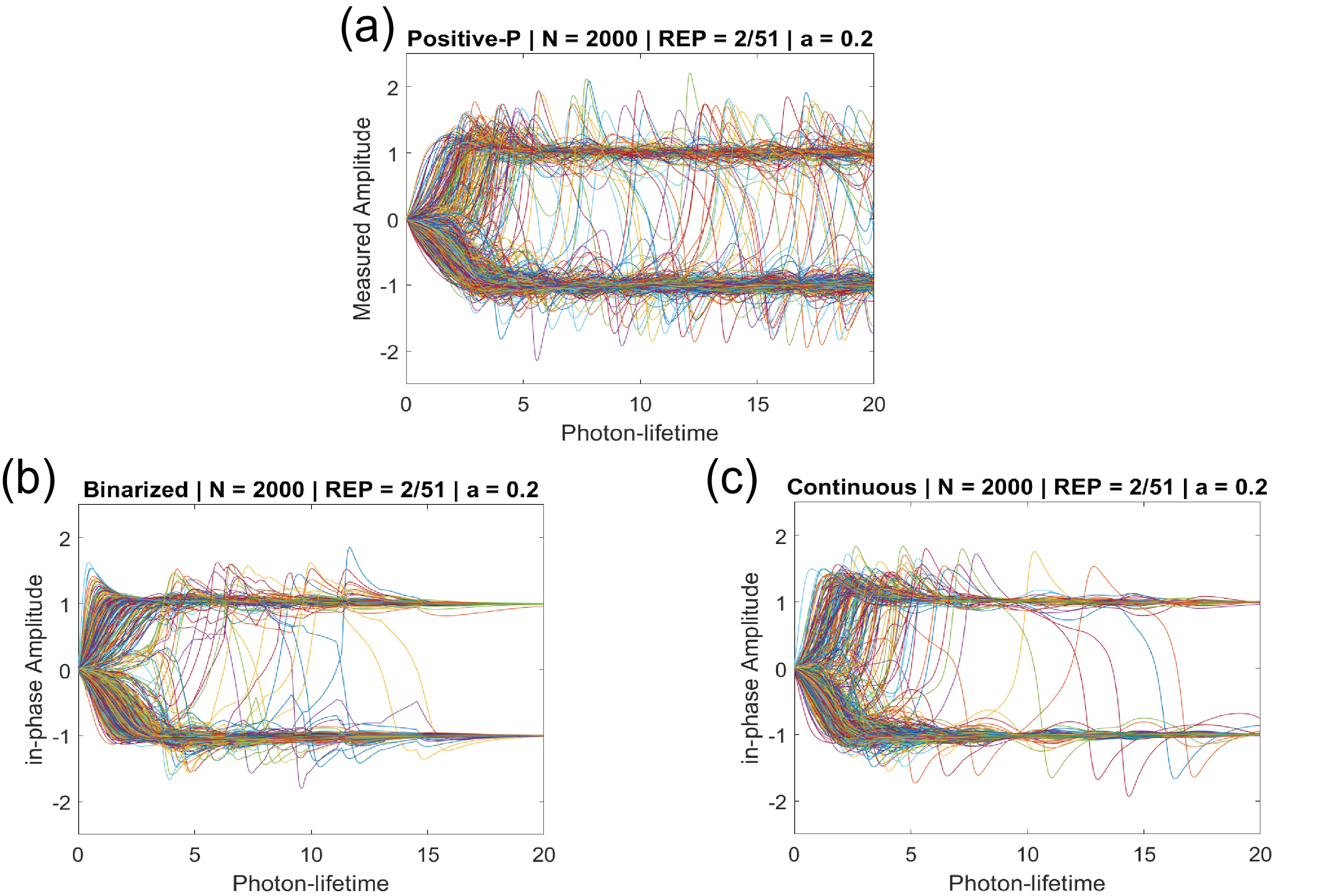}
\caption{\label{amps} \textbf{Amplitude evolution of CAC-CIM (Positive-$P$) and CAC-MFZs with continuous and binarized local fields.} \textbf{(a)}, \textbf{(b)} and \textbf{(c)} indicates the measured amplitude $\tilde{\mu}_r$ of CAC-CIM (Positive-$P$), amplitude $c_r$ evolution of CAC-MFZ with continuous and binarized local fields introduced in eq. (\ref{localfieldMFZ}) and eq. (\ref{localfieldMFZBN}) respectively. A slightly more chaotic behavior can be seen in the continuous case. {The indicated amplitudes for each model come from the second alternative minimization process in Algorithm} \ref{algo} {under the same problem instance.} The system size was set as $N = 2000$ while the compression and the sparseness were 0.6 and 0.2. 
}
\end{figure}

\begin{figure*}[!ht]
\includegraphics[width=140mm]{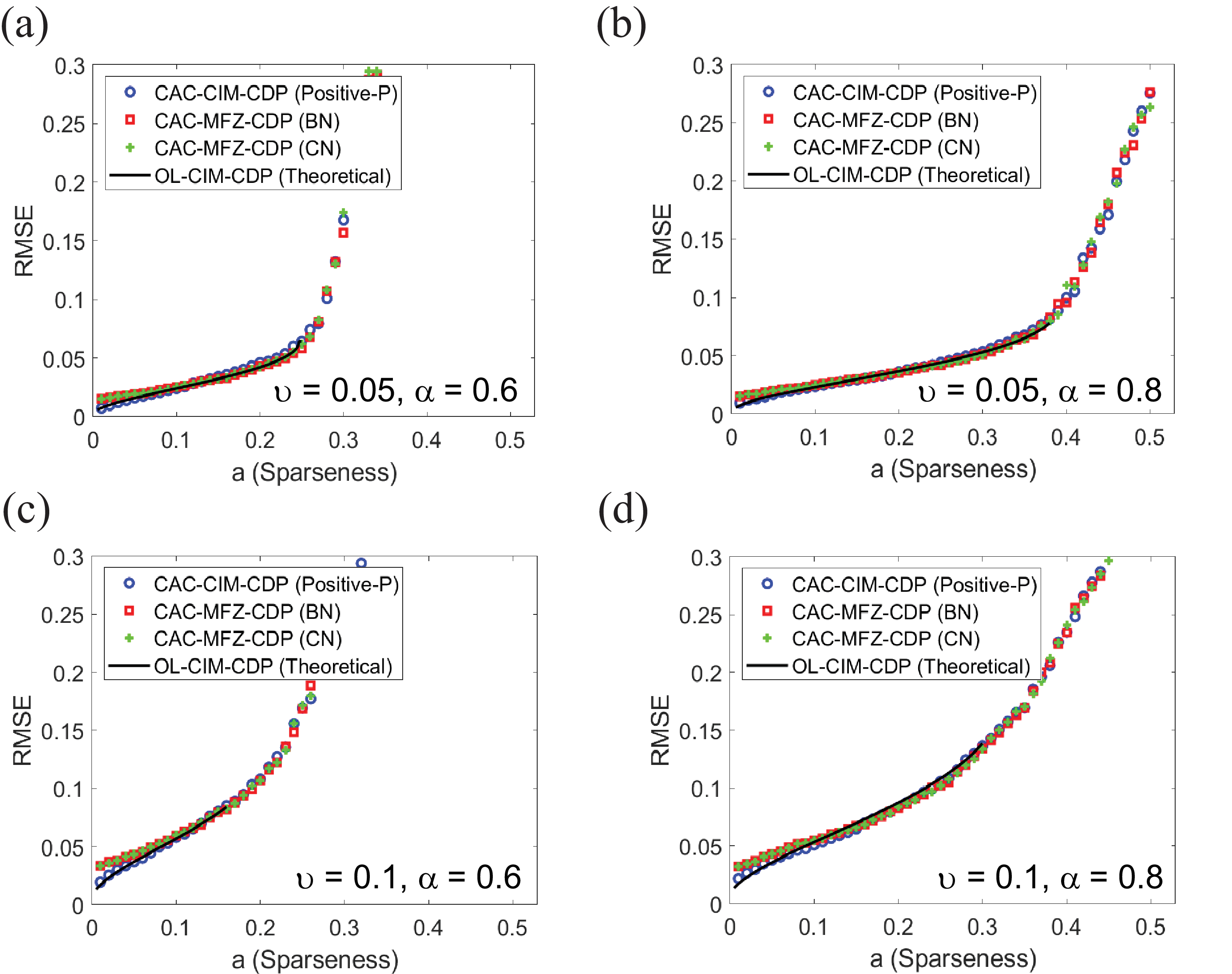}
\caption{\label{noisygraph} \textbf{CAC-MFZ-CDP's average RMSE compared to the theoretical limit of L0RBCS, when observation noise exists.} \textbf{(a)} and \textbf{(b)} indicates the average performance for $N = 2000$ system where $\alpha = 0.6$ and $\alpha = 0.8$ respectively for $\nu = 0.05$. \textbf{(c)} and \textbf{(d)} states the average performance for $\nu = 0.1$. 
$\eta_{init} = 0.8$ was used for CAC-CIM-CDP and CAC-MFZ-CDP. \textbf{(a)} and \textbf{(b)} $\eta_{end}$ was set to 0.18 for all models. \textbf{(c)} and \textbf{(d)} $\eta_{end}$ was set to 0.35 for all models.}
\end{figure*}

\subsection*{Comparison of ground state predictions with alternating minimization} \label{noisy}

In this section, the ability of CAC-MFZ-CDP to find the ground state is compared to L0RBCS's theoretical limit. For a detailed explanation of the non-equilibrium statistical mechanics method to derive the macroscopic parameter equations, see \cite{Aonishi}. 

Here we generate random samples of the observation matrix, source signal, and true support vector in order to compare the solutions of the models with the ground state predicted by statistical mechanics. System size $N$ was set to $2000$ with varying $a, \alpha$ and $\nu$ (standard deviation of the observation noise) values. $\eta_{init}$ was set to 0.8 for CAC-CIM-CDP and CAC-MFZ-CIM. For $\nu = 0.05$ (\ref{noisygraph}a and \ref{noisygraph}b), $\eta_{end}$ was $0.18$ where $\eta_{end} = 0.35$ for $\nu = 0.1$ (\ref{noisygraph}c and \ref{noisygraph}d). 

In the $y$-axis of Fig. \ref{noisygraph} averaged root-mean-square-error (RMSE) (defined as $\sqrt{1/N \sum_{r=1}^N \left(R_r\sigma_r - x_r\xi_r\right)^2}$) is indicated. Here $x_r$ and $\xi_r$ are the correct $N$-dimensional signal and support vectors (see Methods). {The black solid line depicts the RMSE predicted by statistical mechanics for L0RBCS defined in eqs.} at its ground state \cite{Aonishi}.
 The red squares represent CAC-MFZ-CDP with the binarized local field $h_r^{BN}$, while the green pluses represent CAC-MFZ-CDP with the continuous local field $h_r$. The blue circles represent the results for CAC-CIM-CDP (Positive-$P$). Results indicate that the performance of both CAC-MFZ-CIM models is similar to CAC-CIM-CDP (Positive-$P$). However, it appears that CAC-MFZ-CDP has slightly worse performance in low $a$ values. {And as can be seen from a comparison of the results between CAC-MFZ-CDP models with $h_r^{BN}$ and $h_r$, the performance is virtually the same.}

\subsection*{Sparse MRI simulations}\label{mri}

Since artificial random data simulations indicate a similar performance to CAC-CIM-CDP, we conducted experiments on MRI data to ascertain whether this similarity exists with real-world data as well.
For each threshold $\eta$ in the images $64\times64$ and $128\times128$, the average RMSE value for 10 simulations is shown in Fig. \ref{mri64}a and Fig. \ref{mri64}b.
The minimum RMSE for $64\times 64$ in terms of LASSO (black line), {CAC-CIM-CDP (Positive-$P$) (dotted blue), CAC-MFZ-CDP with the binarized local field $h_r^{BN}$ (red) and CAC-MFZ-CDP with the continuous local field $h_r$ (green) can be indicated by 0.0292, 0.0182, 0.0176 and 0.0168, respectively.} Fig.\ref{mriimages} {indicates the corresponding reconstructions for the case $64\times 64$ pixels. Furthermore,} Fig.\ref{mriimages} {illustrates the difference in pixel identification between the two models when compared to the original resized image for case $64\times 64$ pixels. However, it is difficult to visually verify the reconstruction accuracy among CAC-MFZ-CDPs with $h_r^{BN}$ and $h_r$. 

For the case of $128\times 128$, the minimum RMSE for each model is 0.0276, 0.0209, 0.0206 and 0.0206, respectively. Here, we do not indicate the reconstruction for $128\times 128$ pixels case because the difference in reconstruction between these CIM models cannot be visually confirmed. According to the RMSE values acquired for various $\eta$, the performance of both the binarized and continuous CAC-MFZ-CDP models is almost the same.  Additionally, both CAC-MFZ-CDP and CAC-CIM-CDP (Positive-$P$) models exhibit relatively similar results.}

MRI data, however, indicates that CAC-MFZ-CDP's average RMSE is higher than CAC-CIM-CDP's in low $\eta$ values such as $\eta < 0.01$. Considering the main difference between CAC-MFZ-CDP and CAC-CIM-CDP is the absence of noise, it is reasonable to assume that quantum noise may boost performance in low $\eta$ ranges for CAC-CIM-CDP.

\begin{figure}[!ht]
\includegraphics[width=140mm]{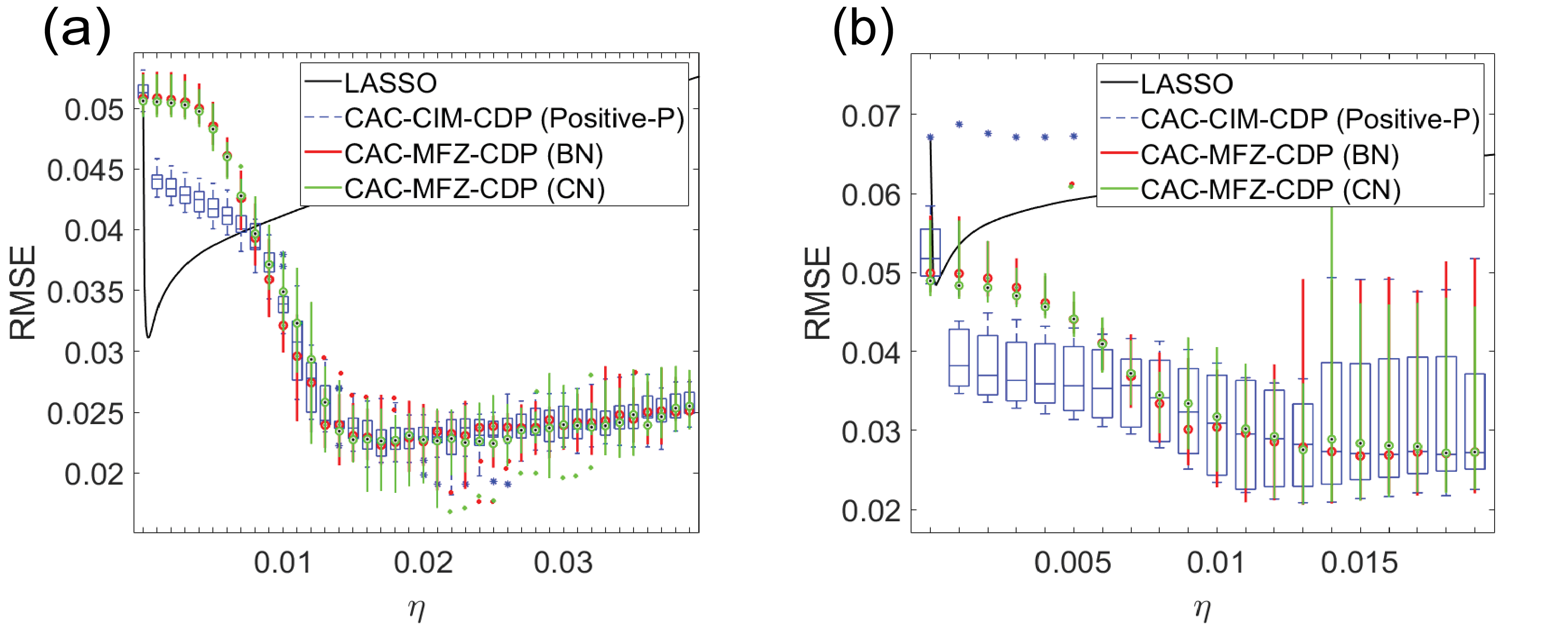}
\caption{\label{mri64} \textbf{Average performance of the models for different-size MRI data when L0-regularization parameter varies}
\textbf{(a)} Performance on $64\times64$. \textbf{(b)} Performance on $128\times128$. The black line indicates LASSO performance. On the blue box plot, CAC-CIM-CDP (Positive-$P$) results are shown, and on the red and green box plots, CAC-MFZ-CDP results with binarized and continuous local fields are shown. Each box plot illustrates the maximum, minimum, 25th percentile (bottom edge), 75th percentile (top edge), and median (central horizontal line) of RMSEs for each model at different threshold values. We display the maximum, minimum, and median of RMSE for CAC-MFZ-CDP. {There are markers indicating outliers.} 
The compression and sparseness for \textbf{(a)} were 0.4 and 0.212 respectively while for \textbf{(b)} were 0.3 and 0.178.}
\end{figure}

\begin{figure*}[!ht]
\includegraphics[width=135mm]{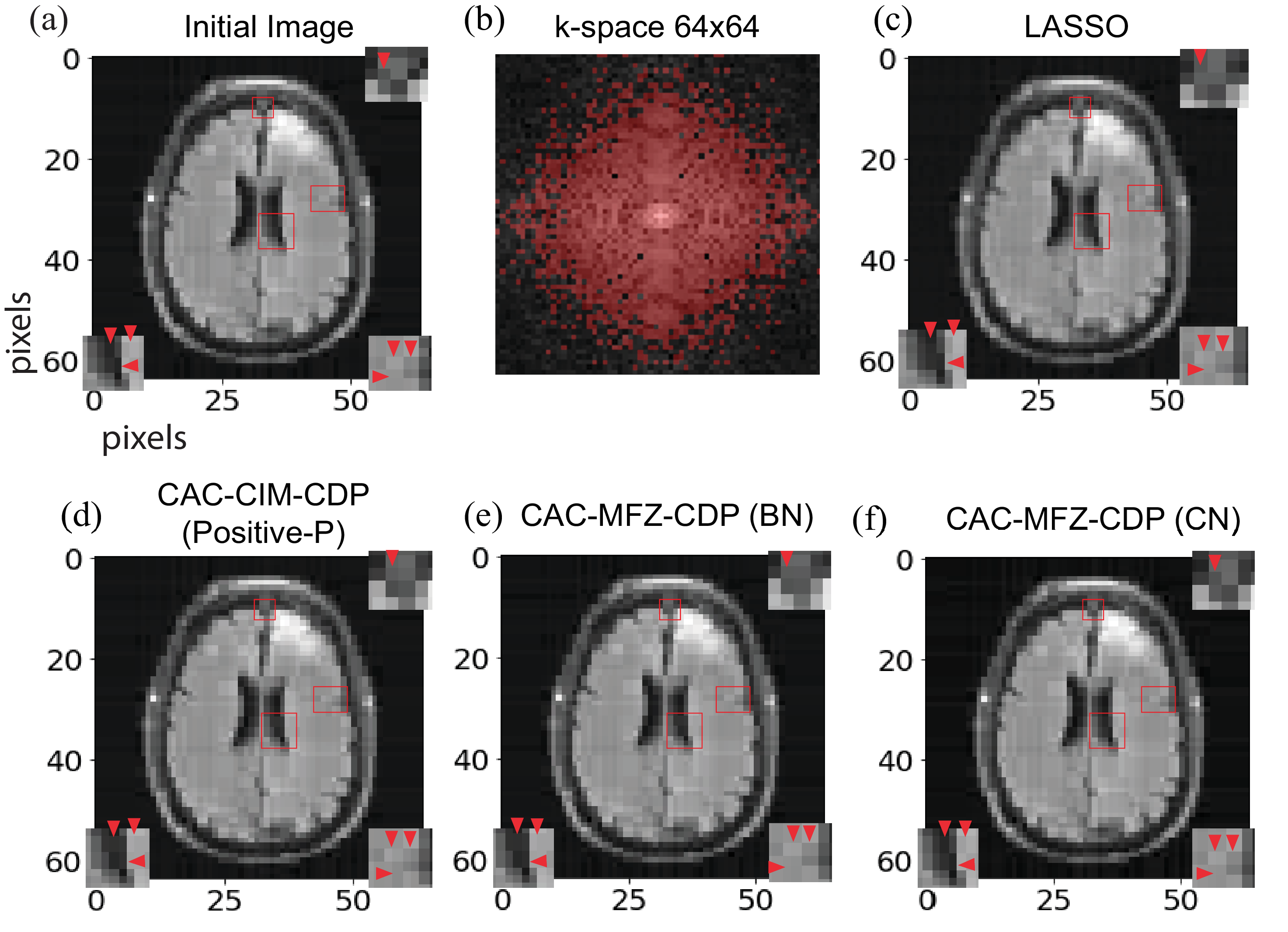}
\caption{\label{mriimages} \textbf{Reconstructed Images for MRI data with $64\times64$.} \textbf{(a)} Initial $64\times 64$ image. Sparseness and compression ratios were 0.212 and 0.4 respectively. \textbf{(b)} {k-space data (grey points) and random undersampled points (red points). \textbf{(c)}, \textbf{(d)}, \textbf{(e)} and \textbf{(f)} correspond to the reconstructions obtained from LASSO, CAC-CIM-CDP (Positive-$P$) and CAC-MFZ-CDPs with continuous and binarized local fields (BN and CN) where RMSE values are 0.0292, 0.0182, 0.0176 and 0.0168 respectively. Pixel-wise differences between the reconstructions are depicted in the enlarged portions of the images. For \textbf{(d)}, \textbf{(e)} and \textbf{(f)} a total of 11 alternating minimization processes were performed. And for \textbf{(c)}, \textbf{(d)}, \textbf{(e)} and \textbf{(f)} $\eta_{init} = \eta_{end}$ was 0.0003, 0.022, 0.025 and 0.022 respectively. }}
\end{figure*}

\section*{Discussion}

\subsection*{Role of quantum noise in CIMs}\label{cimnoise}

CAC-CIM and MFZ-CIM differ predominantly in their presence and absence of quantum noise. 
As the saturation parameter $g^2 = 10^{-7}$ is taken into account in the CAC-CIM, the model only introduces a subtle amount of noise to the system.

There are, however, some disadvantages associated with the presence of quantum noise.
Multiple studies have demonstrated that increasing $g^2$ (aka decreasing the number of photons per pulse, which increases quantum noise) decreases the success probability of the CIM \cite{Inui, Inui2022, gunathilaka2023}. It is mainly due to the fact that when indirect homodyne measurements are carried out, those signals are easily buried in the quantum noise in the CIM \cite{inui2024skewgaussian}. {Essentially, this is due to the fact that, at around $g^2 = 10^{-2}$, when $g^2$ is continued to increase, the amplitude fluctuation becomes larger than the target amplitude $\tau$} (eq. (\ref{GaussianEC5}) and $\tau = 1$), {and with such large fluctuations, $e_r$ will keep decreasing (see Supporting information), resulting in no effective feedback} \cite{Inui2022}. 

{Initially, we supposed that the introduction of subtle noise would increase the success probability and help CAC-CIM reach the theoretical limit of L0RBCS in small $a$ regions with large observation noise ($\nu = 0.1$) (see Figs.} \ref{noisygraph}c and \ref{noisygraph}d {which indicates slightly low RMSE with CAC-CIM for $a < 0.05$ than with MFZ-CIM). However, further numerical simulations revealed that quantum noise may not be the cause of this deviation, which rather may be caused by parameter optimization (see Supporting information).}

As of right now, quantum noise appears to be a disadvantage in CIM. Nevertheless, a more comprehensive study of this topic may contribute to a deeper understanding of the CIM.

\subsection*{Advantage of Mean-Field CIM for large-scale optimization}\label{largescale}

In comparison to CAC-CIM (eq. (\ref{ppGACsCIM1})-(\ref{localfieldGACS})), it is evident that MFZ-CIM (eq. (\ref{mfeq1})-(\ref{localfieldMFZBN})) has a much simpler formulation. 
However, the MFZ-CIM is a heuristic approach. Since real-world optimization problems are large-scale, a less computationally intensive solver is essential. So MFZ-CIM is suitable for such large-scale COPs. 
Furthermore, it has been claimed that in MF-CIM the success rate of finding a solution has been higher \cite{Leleuscaling}. 

\subsection*{Difference in three fields}\label{cnbnlocals}

We compared the performance of these three models using artificial random data with an additional experiment shown in Fig. \ref{contvsbinrmse}, in order to determine if there are any significant performance differences between them.
 In Fig. \ref{contvsbinrmse}a, the RMSE for every alternative minimization procedure is shown as boxplots for CAC-CIM-CDP (Positive-$P$), binarized local field CAC-MFZ-CDP, and continuous local field CAC-MFZ-CDP. Each box plot illustrates the maximum, minimum, 25th percentile (bottom edge), 75th percentile (top edge), and median (central horizontal line) of RMSEs for each model at different alternative minimization steps. From Fig. \ref{contvsbinrmse}a, it can be concluded that these three methods have relatively similar performance. Furthermore as shown in Fig. \ref{contvsbinrmse}b, the Hamming loss (determines how accurate the prediction of the support) was calculated as $Hamming \ Loss \ = 1/N \left( \sum_{r = 0}^{N} \lvert\sigma_r - \xi_r\rvert \right)$, the CAC-MFZ-CDP method had slightly better accuracy than CAC-CIM-CDP at predicting the support in at certain alternative minimization steps (e.g. step 30 - step 40) and the support estimation accuracies of CAC-MFZ-CDPs with continuous local fields and discrete local fields are almost the same as each other. Therefore, CAC-MFZ-CDP slightly outperforms the CAC-CIM-CDP, which is consistent with our previous reports \cite{gunathilakameanfield}.
 And, Figs. \ref{contvsbinrmse}a and \ref{contvsbinrmse}b also support that there is almost no difference in performance between CAC-MFZ-CDPs with continuous local fields and discrete local fields.

\begin{figure}[!ht]
\includegraphics[width=140mm]{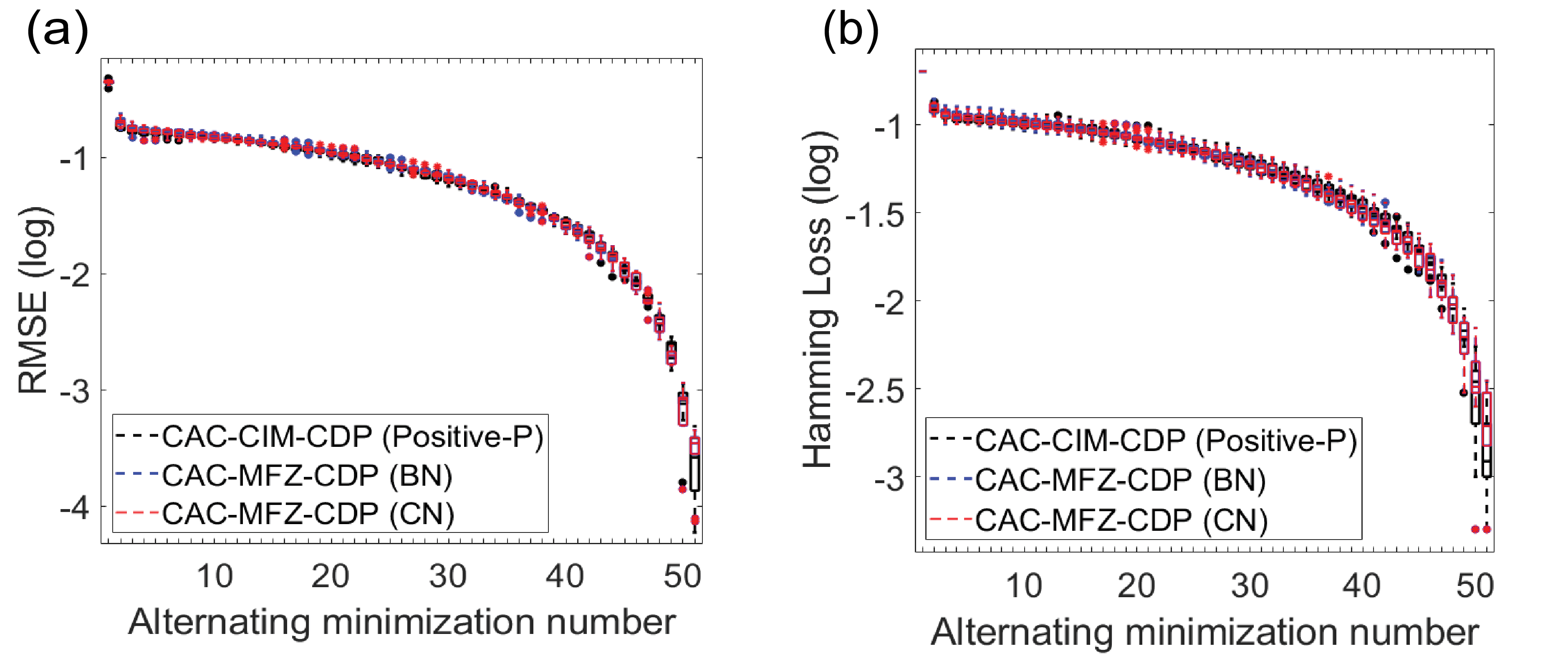}
\caption{\label{contvsbinrmse} \textbf{Performance comparison between continuous and binarized local fields at each alternative minimization step.} \textbf{(a)} Log average RMSE values for 10 random datasets for each alternative minimization step. CAC-CIM-CDP (Positive-$P$), CAC-MFZ-CDP (BN) and CAC-MFZ-CDP (CN) results are indicated in black, blue and red boxplots respectively. \textbf{(b)} Log average Hamming loss values for 10 random datasets for each alternative minimization step which calculates the average support estimation accuracy.}
\end{figure}

\subsection*{Advantages of discrete local fields}\label{discreteadvans}

One of the principal advantages of discrete local fields is their simplicity when it comes to hardware implementations such as FPGAs. 
The most expensive process in the calculation of the CIM models is the multiplication of the $N\times N$ coupling matrix with the $N$-dimensional amplitude vector for calculating the local field. 
By comparing discrete variable local fields with continuous variable local fields in our formulation, one can see that the complexity and memory requirements of the circuit can be substantially reduced with the discrete local field, making it possible to increase the degree of parallelism for the multiplication operation. Furthermore, all the results in this paper, including Fig. \ref{contvsbinrmse}, show that CAC-MFZ-CDPs with continuous local fields and discrete local fields perform almost identically. Thus, using discrete local fields, it is possible to realize highly parallel implementation of large-scale systems into digital circuits such as FPGAs without compromising the performance.

\subsection*{Future work}

Taking into account the paper's goal to compare CAC-MFZ-CDP performance with CAC-CIM-CDP (low computational cost model vs. high computational cost model), the acquired results indicate that both models have a reasonably similar performance on L0RBCS. Therefore, CAC-MFZ-CDP is a relatively low computational cost alternative to complicated CIM SDEs for L0RBCS. With this, it should make it relatively easy to implement on digital hardware such as FPGAs. To our knowledge as of yet, no CIM models have been implemented on FPGAs with Zeeman terms, or CIM models that solve real-world optimization problems. It is our belief that the CAC-MFZ-CDP model proposed in this paper will bring CIMs one step closer to being used to solve real-world optimization problems in real life. 

However, it is evident that how effective the CAC algorithm is for large-scale COPs is still unknown. Further investigation is needed in this context on how quantum noise in CIM and CAC's chaotic behavior affects the performance of the given problem. Even though it is computationally costly for large-scale COPs to compare every spin-flip and decide the optimal rather than just taking the final spin configuration as the estimation after time development, an investigation is needed in order to determine which is better due to the difference between amplitude evolution of the models.

Furthermore, parameter optimization has been shown to improve the performance of these models in certain regions (see Supporting information). 
It is pertinent to note that using grid-search methods to find optimal parameters may not be the best / most effective method when attempting to solve large-scale optimization problems. 
Methods such as Bayesian optimization leverage historical evaluations to make informed decisions \cite{bayseanopt}. It utilizes past data to minimize search time and enhance model performance by identifying optimal configurations of hyperparameters. Compared to traditional methods such as grid search, these methods are generally considered to be more effective in scenarios with high-dimensional hyperparameter spaces. For future implementations, it may be beneficial to consider using such techniques in order to evaluate how much of an accuracy improvement could be achieved.

\section*{Conclusion}

In this paper, we have introduced CAC-MFZ-CDP, a physics-inspired heuristic model for solving L0RBCS problems. In contrast to CAC-CIM-CDP, a more physically accurate model for L0RBCS, CAC-MFZ-CIM displays similar performance in both artificial random data and real-world MRI data. In the new model, we have introduced a binarized local field along with a continuous local field that is tailored for future digital hardware implementations. With its simplicity, CAC-MFZ-CIM overcomes the computational cost of CAC-CIM-CDP and paves the way for FPGA-based digital hardware implementations.

\section*{Supporting information}

\appendix
\subsection*{Derivation of the injection field for CAC-MFZ-CDP}

In the case of L0RBCS, consider the Hamiltonian as follows.

\begin{equation}
\label{eq1}
    \mathbcal{H} = \sum_{r<r'}^{N}\sum_{k = 1}^{M} A_{r}^{k}A_{r'}^{k}R_{r}R_{r'}\sigma_{r}\sigma_{r'} - \sum_{r=1}^{N}\sum_{k =1}^{M} y^{k}A_{r}^{k}R_{r}\sigma_{r} + {\lambda} \sum_{r = 1}^{N} \sigma_r .
\end{equation}

\noindent Then considering Ising spins as $s_r = \pm 1$, Hamiltonian is converted using a quadratic unconstrained binary optimization (QUBO) problem conversion.

\begin{equation}
\begin{multlined}
\label{eq2}
    \mathbcal{H} = \sum_{r<r'}^{N}\sum_{k = 1}^{M} A_{r}^{k}A_{r'}^{k}R_{r}R_{r'}\left(\dfrac{s_{r}+1}{2}\right)\left(\dfrac{s_{r'}+1}{2}\right)\\ - \sum_{r=1}^{N}\sum_{k =1}^{M} y^{k}A_{r}^{k}R_{r}\left(\dfrac{s_{r}+1}{2}\right) + {\lambda} \sum_{r = 1}^{N} \left(\dfrac{s_{r}+1}{2}\right) .
\end{multlined}
\end{equation}

\noindent Accordingly, if we take the derivative with respect to $s_{r}$ into account, we can write the injection field in the following way.

\begin{equation}
\begin{multlined}
\label{eq3}
    \left(\dfrac{dc_{r}}{dt}\right)_{inj,r} = -\sum_{r<r'}^{N}\sum_{k = 1}^{M} \dfrac{1}{2} A_{r}^{k}A_{r'}^{k}R_{r}R_{r'}\left(\dfrac{s_{r'}+1}{2}\right)\\ + \dfrac{1}{2}\sum_{r=1}^{N}\sum_{k =1}^{M} y^{k}A_{r}^{k}R_{r} - \dfrac{\lambda}{2} .
\end{multlined}
\end{equation}

\noindent By replacing $1$ with $\sqrt{\tau}$, $s_r$ with $c_r$, and the regularization parameter using $\eta$, we get the injection field for continuous variables.

\begin{equation}
\begin{multlined}
\label{eq4}
 \left(\dfrac{dc_{r}}{dt}\right)_{inj,r}  = -{\sum_{r' = 1 (\neq r)}^{N}\sum_{k = 1}^{M}}\dfrac{1}{2} A_{r}^{k}A_{r'}^{k}R_{r'}\dfrac{1}{2}\left(c_{r'} + \sqrt{\tau} \right) {+} \sum_{k = 1}^{M} \dfrac{\sqrt{\tau}}{2}{A_{r}^{k}y^{k}}\\ - \dfrac{\sqrt{\tau}\eta^2}{4} .
\end{multlined}
\end{equation}

\noindent We replace $\left(c_{r'} + \sqrt{\tau} \right)/2$ with $\sigma_{r'}$ in order to transform eq. (\ref{eq4}) into a binarized formulation.
As a result, we have the following injection field.

\begin{equation}
\begin{multlined}
\label{eq5}
 \left(\dfrac{dc_{r}}{dt}\right)_{inj,r}  = -{\sum_{r' = 1 (\neq r)}^{N}\sum_{k = 1}^{M}}\dfrac{1}{2} A_{r}^{k}A_{r'}^{k}R_{r'}\sigma_{r'} {+} \sum_{k = 1}^{M} \dfrac{\sqrt{\tau}}{2}{A_{r}^{k}y^{k}}\\ - \dfrac{\sqrt{\tau}\eta^2}{4} .
\end{multlined}
\end{equation}

\noindent If necessary, the $1/2$ in the injection field can be ignored since it is a result of replacing $\sigma_r$ with $(c_r+1)/2$ before taking the derivative with respect to $c_{r}$. Replacing $\sigma_r$ with $(c_r+1)/2$ after taking the derivative with respect to $\sigma_r$ will not include a $1/2$. 

\subsection*{{Error amplitude when $g^2$ becomes larger}}\label{erramps}

This phenomenon is addressed in detail in Ref. \cite{Inui2022} where the performance of low-photon CIM was examined. In Fig. \ref{effeedfig} $y$-axis indicates the log-values of the error amplitude where $x$-axis indicates the photon-lifetime. Fig. \ref{effeedfig}a and Fig. \ref{effeedfig}b corresponds to $g^2 = 10^{-7}$ and $10^{-1}$ respectively for the second alternating minimization process where Fig. \ref{effeedfig}c and Fig. \ref{effeedfig}d corresponds to $g^2 = 10^{-7}$ and $10^{-1}$ respectively for the twentieth one. It is evident from Fig. \ref{effeedfig} that when $g^2$ becomes larger, $e_r$ keeps decreasing resulting in no effective feedback.

\begin{figure}[!ht]
\includegraphics[width=135mm]{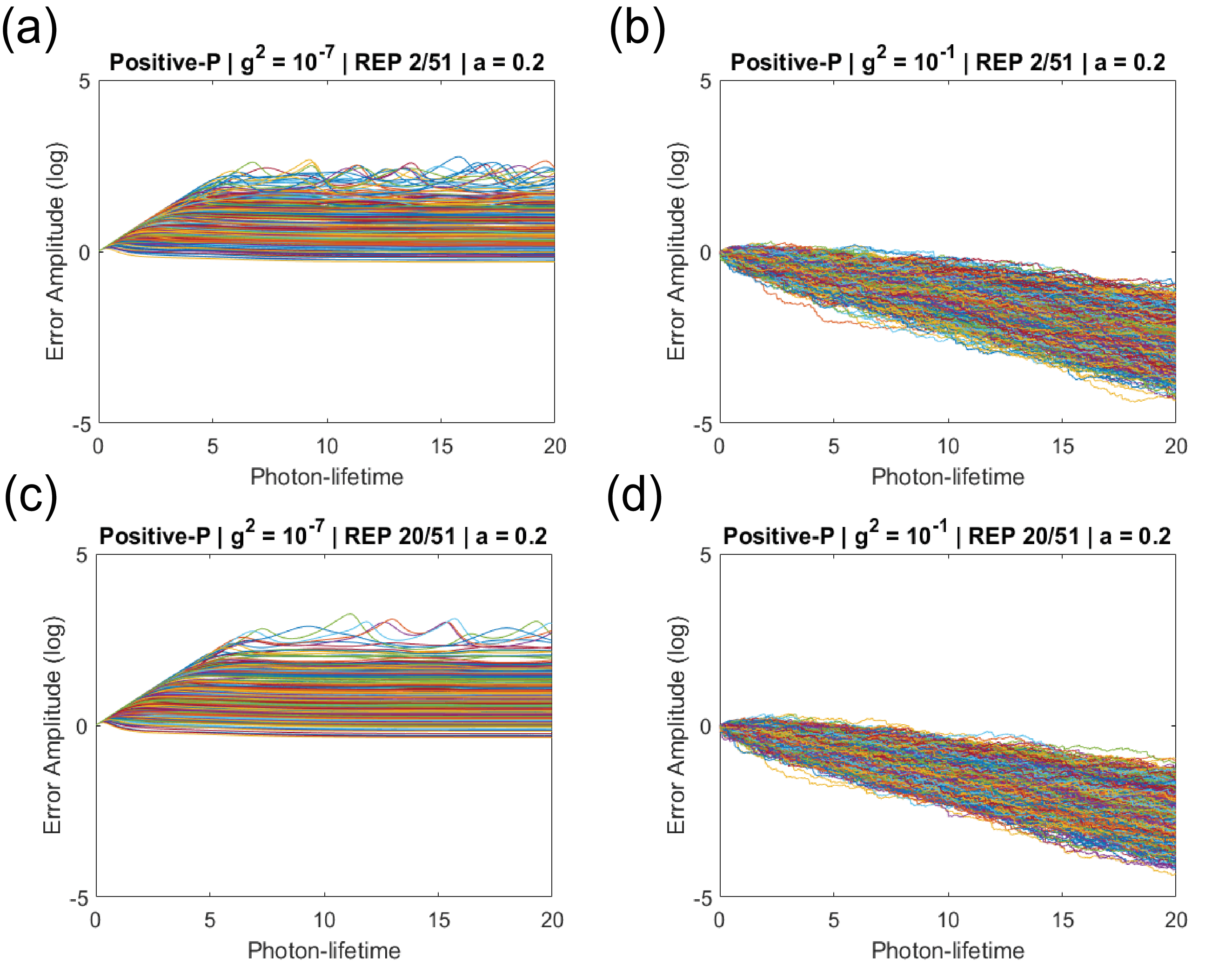}
\caption{\label{effeedfig} \textbf{Error amplitude ($e_r$) evolution of CAC-CIM (Positive-$P$) when $g^2 = 10^{-7}$ and $g^2 = 10^{-1}$.} \textbf{(a)} and \textbf{(c)} indicate the error amplitude $e_r$ (in log values) of CAC-CIM (Positive-$P$) when $g^2 = 10^{-7}$ in the second and twentieth alternative minimization process respectively in Algorithm \ref{algo}. \textbf{(b)} and \textbf{(d)} indicate the $e_r$ of CAC-CIM (Positive-$P$) when $g^2 = 10^{-1}$ in the second and twentieth alternative minimization process.
 {The results are under the same problem instance.} The system size was set as $N = 2000$ while the compression and the sparseness were 0.6 and 0.2. 
}
\end{figure}

\subsection*{{Performance discrepancy around $a < 0.05$}}\label{effeed}

During the analysis of Fig. \ref{noisygraph}, it was clear that both the CAC-MFZ-CDP (BN) and CAC-MFZ-CDP (CN) performed slightly worse than the CAC-CIM-CDP (Positive-$P$) model in the area of $a < 0.05$. In the Discussion, we explained that one of the major differences between CAC-CIM and MFZ-CIM is the absence of quantum noise in MFZ-CIM. Initially, this performance discrepancy was thought to be caused by the effect of quantum noise in the CAC-CIM and that this noise allows the CAC-CIM to perform better compared to CAC-MFZ-CDP (BN) and CAC-MFZ-CDP (CN). Further calculations, however, have demonstrated that this may not be the case.

\begin{figure}[!ht]
\includegraphics[width=135mm]{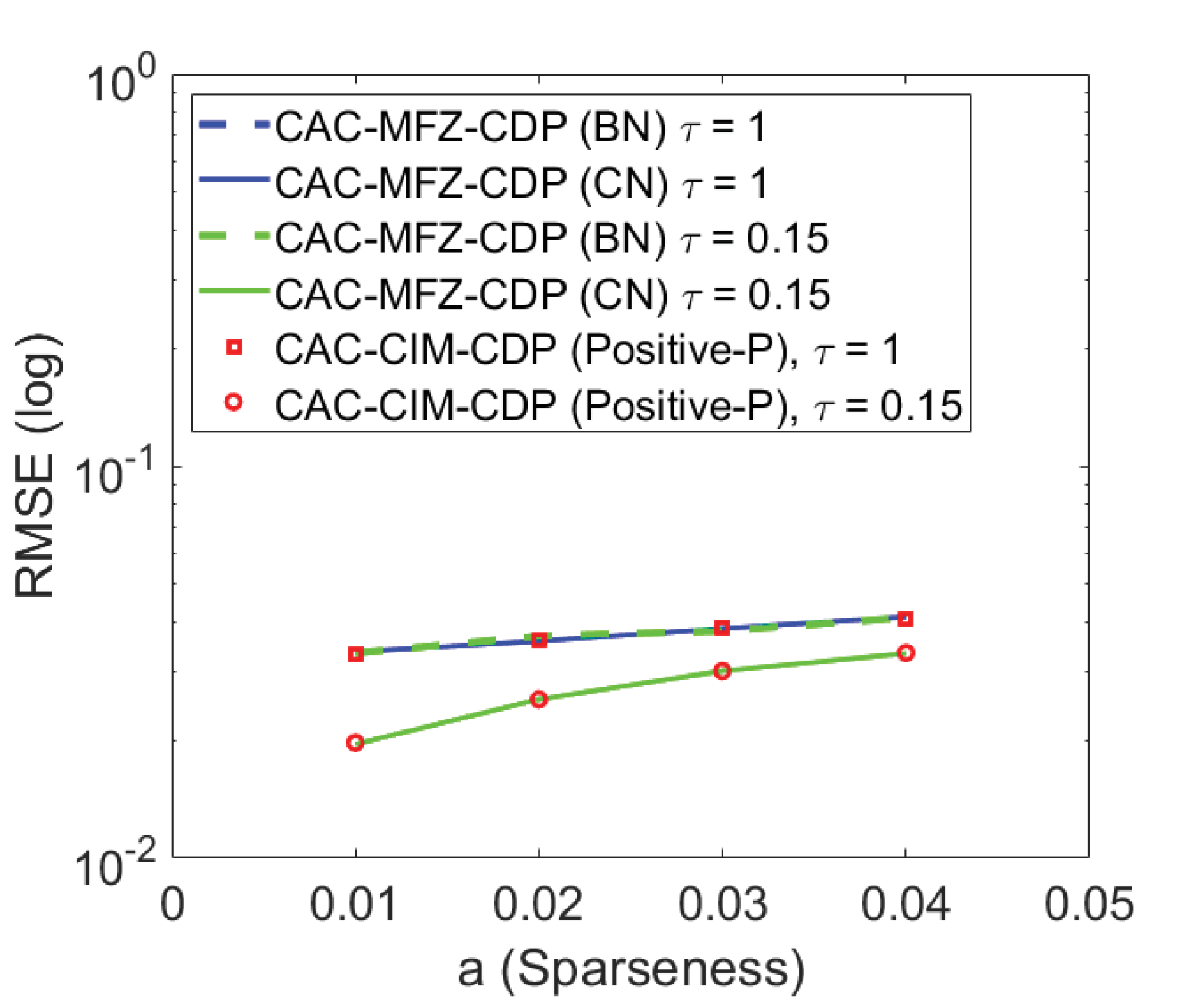}
\caption{\label{gnorm} \textbf{Performance difference when sparseness and $\tau$ changes for CAC-CIM-CDP (Positive-$P$).} The $y$-axis indicates the log-RMSE values acquired when sparseness and $\tau$ is changed for each model. Dashed blue and green lines indicate the CAC-MFZ-CDP (BN) with $\tau = 1$ and $\tau = 0.15$ respectively. The solid blue and green lines indicate CAC-MFZ-CDP (CN) for $\tau = 1$ and $\tau = 0.15$ respectively. Red square and circle indicate the CAC-CIM-CDP (Positive-$P$) with $\tau = 1$ and $\tau = 0.15$ respectively. Here $\nu = 0.1$, $\alpha = 0.6$ and $N = 2000$.
}
\end{figure}

The analysis was conducted by considering CAC-CIM-CDP (Positive-$P$) eq. (\ref{ppGACsCIM1}) - (\ref{ppGACsCIM3}) using $g^2 = 10^{-7}$. 
In Fig. \ref{gnorm}, dashed blue and green lines indicate the CAC-MFZ-CDP (BN) with $\tau = 1$ and $\tau = 0.15$ respectively. The solid blue and green lines indicate CAC-MFZ-CDP (CN) for $\tau = 1$ and $\tau = 0.15$ respectively. 
In this case, we consider the first term of eq. (\ref{mfeq1}) as $(-1+p-j-c_r^2)c_r$ instead of $(-1+p-c_r^2)c_r$ to match the term of eq. (\ref{ppGACsCIM1}) for CAC-CIM-CDP (Positive-$P$).
The red square and circle indicate the CAC-CIM-CDP (Positive-$P$) with $\tau = 1$ and $\tau = 0.15$ respectively. As it is clear from the results, when $\tau = 0.15$ is used, the performance of the CAC-CIM-CDP (Positive-$P$) performance becomes almost identical to CAC-MFZ-CDP (CN). However, CAC-MFZ-CDP (BN) does not show a difference in performance with the change in $\tau$. Since the continuous and binarized models have different compositions, this is to be expected. Consequently, the continuous model has more sensitive parameters than the binarized model, and the discrepancy around $a < 0.05$ can be explained by parameter optimization.

\section*{Acknowledgments}
This work is supported by NTT Research Inc. And the Authors acknowledge the support of the NSF CIM Expedition award (CCF-1918549).



%
%
%



\end{document}